\documentclass[10pt]{article}

\usepackage{amsmath}
\usepackage{amssymb}
\usepackage{amsthm}
\usepackage{bbm,bm}
\usepackage{color}
\usepackage{subfigure}
\usepackage{graphicx}
\usepackage{graphics}
\usepackage[pdftex,pagebackref,letterpaper=true,colorlinks=true,pdfpagemode=none,urlcolor=blue,linkcolor=blue,
citecolor=blue,pdfstartview=FitH,plainpages=true]{hyperref} 
\usepackage[comma,authoryear]{natbib}

\newcommand{\cov}[0]{\mathrm{cov}}
\newcommand{\var}[0]{\mathrm{var}}
\newcommand{\pr}[0]{\mathbb{P}}
\newcommand{\R}[0]{\mathbb{R}}

\newcommand{\E}[0]{\mathbb{E}}
\newcommand{\tr}[0]{\mathrm{tr}}

\def \bbE {\mathbb{E}}
\def \bbF {\mathbb{F}}

\def \bbP {\mathbb{P}}
\def \bbR {\mathbb{R}}

\renewcommand{\hat}{\widehat}
\renewcommand{\tilde}{\widetilde}

\theoremstyle{plain}
\newtheorem{theorem}{Theorem}[section]
\newtheorem{lemma}[theorem]{Lemma}
\newtheorem{proposition}[theorem]{Proposition}
\newtheorem{corollary}[theorem]{Corollary}

\theoremstyle{definition}

\newtheorem{example}{Example}[section]
\newtheorem{assumption}{Assumption}[section]

\theoremstyle{remark}
\newtheorem{remark}{Remark}

\numberwithin{equation}{section}

\begin{document}

\title{High Dimensional Analysis of Variance in Multivariate Linear Regression}

\author{Zhipeng Lou$^1$, Xianyang Zhang$^2$ and Wei Biao Wu$^3$}
\footnotetext[1]{Department of Operations Research and Financial Engineering, Princeton, NJ 08544.}
\footnotetext[2]{Department of Statistics, Texas A\&M University, College Station, TX 77843.}
\footnotetext[3]{Department of Statistics, University of Chicago, Chicago, IL, 60637.}
\date{\today}
\maketitle


\begin{abstract}
In this paper, we develop a systematic theory for high dimensional analysis of variance in multivariate linear regression, where the dimension and the number of coefficients can both grow with the sample size. We propose a new \emph{U}~type test statistic to test linear hypotheses and establish a high dimensional Gaussian approximation result under fairly mild moment assumptions. Our general framework and theory can be applied to deal with the classical one-way multivariate ANOVA and the nonparametric one-way MANOVA in high dimensions. To implement the test procedure in practice, we introduce a sample-splitting based estimator of the second moment of the error covariance and discuss its properties. A simulation study shows that our proposed test outperforms some existing tests in various settings.

\end{abstract}

\section{Introduction}
In statistical inference of multivariate linear regression, a fundamental problem is to investigate the relationships between the covariates and the responses. In this article, we aim to test whether a given set of covariates are associated with the responses by multivariate analysis of variance (MANOVA). To fix the idea, we build the multivariate linear regression model with $p$ predictors as
\begin{align}
\label{model1}
    Y_{i} = B^{\top} X_{i} + V_{i} \enspace (i = 1, \ldots, n), 
\end{align}
where $Y_{i} = (Y_{i1}, \ldots, Y_{id})^{\top}$ and $X_{i} = (X_{i1}, \ldots, X_{ip})^{\top}$ are respectively the response vector and the predictor vector respectively for the~\emph{i}th sample, $B^{\top} = (B_{1}, \ldots, B_{p})$ is the unknown coefficient matrix with $B_{k} \in \mathbb{R}^{d}$ consisting of coefficients on the~\emph{k}th covariate, and the innovation vectors $V_{1}, \ldots, V_{n} \in \mathbb{R}^{d}$ are independent and identically distributed random vectors with $\mathbb{E} (V_{1}) = 0$ and $\cov(V_{1}) = \Sigma$. The first element of $X_{i}$ can be set to be 1 to reflect an intercept term. Equivalently we can write~\eqref{model1} in compact matrix form as
\begin{align}
\label{model matrix}
    Y = XB + V,
\end{align}
where $Y = (Y_{1}, \ldots, Y_{n})^{\top}$, $X = (X_{1}, \ldots, X_{n})^{\top}$ and $V = (V_{1}, \ldots, V_{n})^{\top}$. Let $\mathcal{C} \in \R^{m\times p}$ be a matrix of rank $m$, where $m \in \{1, \ldots, p\}$. We are interested in testing a collection of linear constraints on the coefficient matrix 
\begin{align}
\label{test}
    H_{0} : \mathcal{C}B = 0 \enspace \mathrm{versus} \enspace H_{1} : \mathcal{C}B \neq 0. 
\end{align}

This testing problem has been extensively studied in the low dimensional setting where both the number of predictors and the dimension of the response are relatively small compared to the sample size. A natural and popular choice is the classical likelihood ratio test when the errors are normally distributed; see Chapter 8 in~\citet{Anderson2003} for a review of theoretical investigations. In recent years, high dimensional data are increasingly encountered in various applications. Over the past decade, there have been tremendous efforts to develop new methodologies and theories for high dimensional regression. The paradigm where $d$ is 1 or small and $p$ can increase with $n$ has received considerable attention, while on the other hand the one where $d$ is very large and $p$ is relatively small has been less studied. The model~\eqref{model matrix} in the latter setting has been applied to a number of research problems involving high-dimensional data types such as DNA sequence data, gene expression microarray data, and imaging data; see for example~\citet{Zapala19430}, \citet{wessel2006generalized} and~\citet{schork2012statistical}. Those related studies typically generate huge amounts of data (responses) that, due to their expense and sophistication, are often collected on a relatively small number of individuals, and investigate how the data can be explained by a certain number of predictor variables such as the ages of individuals assayed, clinical diagnoses, strain memberships, cell line types, or genotype information~\citep{Zapala19430}. Owing to inappropriateness of applying the standard MANOVA strategy and shortage of high-dimensional MANOVA theory, biological researchers often considered some form of data reduction such as cluster analysis and factor analysis, which can suffer from many problems, as pointed out by~\citet{schork2012statistical}. In the works~\citet{Zapala19430, schork2012statistical}, the authors incorporated a distance matrix to modify the standard MANOVA, but they commented that there is very little published material that can be used to guide a researcher as to which distance measure is the most appropriate for a given situation. Motivated by these real-world applications, we aim to develop a general methodology for high dimensional MANOVA and lay a theoretical foundation for assessing statistical significance.

The testing problem~\eqref{test} for model~\eqref{model matrix} is closely related to a group of high dimensional hypothesis tests. Two-sample mean test, for testing $H_{0}: \mu_{1} = \mu_{2}$ where $\mu_{1} \in \mathbb{R}^{d}$ and $\mu_{2} \in \mathbb{R}^{d}$ are mean vectors of two different populations, is a special case with $p = 2$, $B = (\mu_{1}, \mu_{2})^{\top}$ and $\mathcal{C} = (1, -1)$. There is a large literature accommodating the Hotelling $T^{2}$ type statistic into the high-dimensional situation where $d$ is large; see for example, \citet{Bai1996}, \citet{Chen2010}, \citet{Srivastava13} among many others. It can be generalized to test the equality of multiple mean vectors in high dimensions. Some notable work includes~\citet{Schott2007}, \citet{Cai2014Xia}, \citet{Hu2017}, \citet{Li2017}, \citet{Zhang2017} and \citet{Zhou2017}. In most existing work, the random samples were assumed to be Gaussian or follow some linear structure as that of~\citet{Bai1996}. In contrast, the testing problem we are concerned is much more general. For one thing, all the aforementioned high dimensional mean test problems can be fitted into our framework, apart from which, we can deal with the more general multivariate linear regression in the presence of an increasing number of predictor variables. For another, we do not assume the Gaussianity or any particular structure of the error vectors $\{V_{i}\}_{i = 1}^{n}$.

Throughout the paper, we assume that $p < n$ and the design matrix $X$ is of full column rank such that $X^{\top} X$ is invertible. The conventional MANOVA test statistic for~\eqref{test} is given by
\begin{align}
\label{defteststatistics}
    Q_{n} = |PY|_{\bbF}^{2} = \sum_{i = 1}^{n} \sum_{j = 1}^{n} P_{ij} Y_{i}^{\top} Y_{j},
\end{align}
where $|\cdot|_{\mathbb{F}}$ stands for the Frobenius norm and 
\begin{equation*}
    P = X(X^{\top} X)^{-1}\mathcal{C}^{\top} \{\mathcal{C}(X^{\top} X)^{-1} \mathcal{C}^{\top}\}^{-1} \mathcal{C}(X^{\top} X)^{-1}X^{\top} = (P_{ij})_{n \times n}
\end{equation*}
is the orthogonal projection matrix onto the column space of the matrix $X(X^{\top} X)^{-1}\mathcal{C}^{\top}$. We shall reject the null hypothesis $H_{0}$ if $Q_{n}$ is larger than some critical value. In the univariate case where $d = 1$, the asymptotic behavior of $Q_{n}$ has been extensively studied in literature; see~\citet{Gotze19991} and~\citet{Gotze2002} for detailed discussions. The validity to perform a test for~\eqref{test} using $Q_{n}$ when $d$ is large has been open for a long time. The first goal of the paper is to provide a solution to this open problem by rigorously establishing a distributional approximation of the traditional MANOVA test statistic when $d$ is allowed to grow with $n$. Our key tool is the Gaussian approximation for degenerate~\emph{U}~type statistics: under fairly mild moment conditions, quadratic functionals of non-Gaussian random vectors can be approximated by those of Gaussian vectors with the same covariance structure. It is worth mentioning that~\citet{Chen2018} established a Gaussian approximation result for high dimensional non-degenerate~\emph{U}~statistics by Stein's method, which can not be applied to the degenerate case here. From a technical point of view, we employ completely different arguments to bound distance between the distribution functions of the test statistic and its Gaussian analogue.

The main contributions of this paper are three-fold. Firstly, we develop a systematic theory for the conventional MANOVA test statistic $Q_{n}$ in the high dimensional setting. More specifically, we shall establish a dichotomy result: $Q_{n}$ can be approximated either by a linear combination of independent chi-squared random variables or by a normal distribution under different conditions; see Theorem~\ref{GArho}. While this reveals the interesting theoretical properties of the test statistics, it causes difficulties in applications as one may not know which asymptotic distribution to use in practice. To overcome this difficulty, as the second main contribution of our paper, we propose using a new~\emph{U}~type test statistic. Using the modified test statistic, such a dichotomy does not appear; see Theorem~\ref{GAtheorem} for the asymptotic result. Thirdly, we will propose a new estimator for the second spectral moment of the covariance matrix via a data-splitting technique. To the best of our knowledge, it is the first work concerning an unbiased and ratio consistent estimator in the multivariate linear regression model.

We now introduce some notation. Let $\mathbb{I}\{\cdot\}$ denote the indicator function. For random variables $X \in \mathbb{R}$ and $Y \in \mathbb{R}$, the Kolmogorov distance is defined by $\rho(X, Y) = \sup_{z \in \bbR}|\bbP(X \leq z) - \bbP(Y \leq z)|$. For $q > 0$, we write $\|X\|_{q} = (\bbE|X|^{q})^{1/q}$ if $\bbE|X|^{q} < \infty$. For two matrices $A = (a_{ij})_{i \le I, j \le J}$ and $B = (b_{ij})_{i \le I, j \le J}$, $A \circ B = (a_{ij} b_{i j})_{i \le I, j \le J}$ denotes their Hardmard product. For any positive integer $m$, we use $I_{m}$ to denote $m \times m$ identity matrix. For two sequences of positive numbers $(a_{n})$ and $(b_{n})$, we write $a_{n} \lesssim b_{n}$ if there exists some constant $C$ such that $a_{n}\leq C b_{n}$ for all large $n$. We use $C, C_{1}, C_{2}, \ldots$ to denote positive constants whose value may vary at different places.

\section{Theoretical results}
\label{sec2}
We start with some notational definitions and basic assumptions. Let $\lambda_{1}(\Sigma) \geq \ldots \geq \lambda_{d}(\Sigma) \geq 0$ denote the eigenvalues of $\Sigma = \cov(V_{1})$ and let $\varsigma = |\Sigma|_{\bbF} = \{\sum_{k = 1}^{d} \lambda_{k}^{2}(\Sigma)\}^{1/2}$. For $q \geq 2$, we define 
\begin{align}
\label{eq_bound_var_U}
    M_{q} = \bbE \left|\frac{V_{1}^{\top}V_{2}}{\varsigma}\right|^{q} \enspace \mathrm{and} \enspace L_{q} = \bbE \left|\frac{V_{1}^{\top}\Sigma V_{1}}{\varsigma^{2}}\right|^{q/2}. 
\end{align}

\begin{assumption}
\label{cond_design}
Recall that $P_{11}, \ldots, P_{nn}$ are diagonal elements of the matrix $P$. Assume that
\begin{align*}
    \frac{1}{m}\sum_{i = 1}^{n} P_{ii}^{2} \to 0 \enspace \mathrm{as} \enspace n \to \infty.
\end{align*}
\end{assumption}

\begin{remark}
\label{Remark_P_matrix}
Assumption~\ref{cond_design} is quite natural and mild for testing~\eqref{test}. For instance, it automatically holds for one sample test of mean vector as $m^{-1} \sum_{i = 1}^{n} P_{ii}^{2} = 1/n$. Additionally, in the context of $K$-sample test, as discussed in Section~\ref{Section_one_way_MANOVA}, Assumption~\ref{cond_design} is satisfied as long as the minimum sample size goes to infinity. More generally, since $\sum_{i = 1}^{n} P_{ii} = m$, a simple sufficient condition for Assumption~\ref{cond_design} would be $\max_{1\leq i\leq n} P_{ii} \to 0$. Further discussions on this condition will be given in Remark~\ref{Remark_cond_balance} and 
Example~\ref{Example_cond_Gaussian}.
\qed
\end{remark}

\subsection{Asymptotic distribution of the conventional MANOVA test statistics}
\label{secV}
Under the null hypothesis $\mathcal{C}B = 0$, $PXB = X(X^{\top} X)^{-1} \mathcal{C}^{\top}\{\mathcal{C}(X^{\top} X)^{-1}\mathcal{C}^{\top}\}^{-1} \mathcal{C}B = 0$ and hence $Q_{n} = |PXB + PV|_{\bbF}^{2} \overset{H_{0}}{=} |PV|_{\bbF}^{2}$, which can be further decomposed as
\begin{align}
\label{eq_decomposition_Q_n}
    Q_{n} \overset{H_{0}}{=} \sum_{i = 1}^{n} \sum_{j = 1}^{n} P_{ij} V_{i}^{\top} V_{j} = \sum_{i = 1}^{n} P_{ii} V_{i}^{\top} V_{i} + \sum_{i = 1}^{n} \sum_{j \neq i} P_{ij} V_{i}^{\top} V_{j} = : D_{n} + Q_{n}^{\star}.
\end{align}
Observe that $D_{n}$ is a weighted sum of i.i.d.~random variables and $Q_{n}^{\star}$ is a second order non-degenerate~\emph{U}-statistic of high dimensional random vectors. These two terms can be differently distributed under the high dimensional setting. More specifically, since $D_{n}$ and $Q_{n}^{\star}$ are uncorrelated, we have $\var(Q_{n}) = \var(D_{n}) + \var(Q_{n}^{\star})$, where 
\begin{align*}
    \var(D_{n}) = \sum_{i = 1}^{n} P_{ii}^{2} \|\mathbb{E}_{0}(V_{1}^{\top} V_{1})\|_{2}^{2} \enspace \mathrm{and} \enspace \var(Q_{n}^{\star}) = 2\left(m - \sum_{i = 1}^{n} P_{ii}^{2}\right) \varsigma^{2},
\end{align*}
where $\mathbb{E}_{0}(V_{1}^{\top} V_{1})=V_{1}^{\top} V_{1}-\mathbb{E}(V_{1}^{\top} V_{1})$.
When the dimension $d$ increases with the sample size $n$, the magnitudes of $\var(D_{n})$ and $\var(Q_{n}^{\star})$ can be quite different for non-Gaussian $\{V_{i}\}_{i = 1}^{n}$; cf.~Example~\ref{example3}. As a consequence, $Q_{n}$ can exhibit different asymptotic null distributions. More precisely, to asymptotically quantify the discrepancy between $\var(D_{n})$ and $\var(Q_{n}^{\star})$, under Assumption~\ref{cond_design}, we define
\begin{align*}
    \Lambda^{2} = \frac{\sum_{i = 1}^{n} P_{ii}^{2} \|\mathbb{E}_{0}(V_{1}^{\top} V_{1})\|_{2}^{2}}{m \varsigma^{2}}.
\end{align*}
Before presenting the distributional theory for $Q_{n}$, we first define its Gaussian analogue. Let $Z_{1}, \ldots, Z_{n}$ be i.i.d.~$N(0, \Sigma)$ Gaussian random vectors and write $Z = (Z_{1}, \ldots, Z_{n})^{\top}$. Then the Gaussian analogue of $Q_{n}$ is defined as the same quadratic functional of $\{Z_{i}\}_{i = 1}^{n}$,
\begin{align}
\label{G0}
    G_{n} = |P Z|_{\bbF}^{2} = \sum_{i = 1}^{n} \sum_{j = 1}^{n} P_{ij} Z_{i}^{\top} Z_{j}. 
\end{align}

\begin{theorem}
\label{GArho}
Let $q = 2 + \delta$, where $0< \delta \leq 1$. Suppose Assumption~\ref{cond_design} holds and 
\begin{align}
\label{cond_Delta_q}
    \Delta_{q} = \frac{\sum_{i = 1}^{n} \sum_{j \neq i} |P_{ij}|^{q}}{m^{q/2}} M_{q} + \frac{\sum_{i = 1}^{n} P_{ii}^{q/2}}{m^{q/2}} L_{q} \to 0. 
\end{align}
\begin{enumerate}
    \item Assume $\Lambda \to 0$. Then, under~\eqref{cond_Delta_q} and the null hypothesis, we have 
    \begin{align*}
        \rho(Q_{n}, G_{n}) \leq C_{1} \Lambda^{2/5} + C_{q} \Delta_{q}^{1/(2q + 1)} + C_{2} \left(\frac{1}{m} \sum_{i = 1}^{n} P_{ii}^{2}\right)^{1/5} \to 0. 
    \end{align*}
    \item Assume $\Lambda \to \infty$ and the Lindeberg condition holds for $W_{i} = \mathbb{E}_{0} (P_{ii}V_{i}^{\top} V_{i})/(\Lambda \varsigma \surd{m})$, that is, $\sum_{i = 1}^{n} \mathbb{E}(W_{i}^{2} \mathbb{I}\{|W_{i}| > \epsilon\}) \to 0$ for any $\epsilon > 0$. Then, under the null hypothesis, we have
    \begin{align}
    \label{Qn_CLT}
        \frac{Q_{n} - m \tr(\Sigma)}{\Lambda \varsigma\surd{m}} \Rightarrow N(0, 1). 
    \end{align}
\end{enumerate}
\end{theorem}

\begin{remark}
\label{Remark_Qn}
Theorem~\ref{GArho} illustrates an interesting dichotomy: the conventional MANOVA test statistic $Q_{n}$ can have one of the two different asymptotic null distributions, depending on the magnitude of the unknown quantity $\Lambda$. This nature of dichotomy poses extra difficulty for utilizing $Q_{n}$ to test~\eqref{test} in practical implementation as we need to predetermine which asymptotic distribution to use. Any subjective choice may lead to unreliable conclusion. To illustrate this, suppose now $\Lambda \to 0$. For $\alpha \in (0, 1)$, let $G_{n}^{-1}(\alpha)$ denote the $(1 - \alpha)$th quantile of $G_{n}$. Based on Theorem~\ref{GArho}, an $\alpha$ level test for~\eqref{test} is given by $\Phi_{0} = \mathbb{I}\{Q_{n} > G_{n}^{-1}(\alpha)\}$. However, if one implements $\Phi_{0}$ under the case where $\Lambda \to \infty$, then the type I error of $\Phi_{0}$ satisfies that $\mathbb{P}(\Phi_{0} = 1 \mid H_{0}) \to 1/2$, which implies that $\Phi_{0}$ in this scenario ($\Lambda \to \infty$) is no better than random guessing. 
\qed
\end{remark}


\begin{remark}
Recently much attention has been paid to studying the dichotomy and similar phase transition phenomenon of the asymptotic distribution of classical tests under the high dimensional setting. For instance, \citet{Xu2019pearson} studied the Pearson's chi-squared test under the scenario where the number of cells can increase with the sample size and demonstrated that the corresponding asymptotic distribution can be either chi-squared or normal. \citet{He2021} derived the phase transition boundaries of several standard likelihood ratio tests on multivariate mean and covariance structures of Gaussian random vectors. In addition to these tests, we suspect similar phenomenon can occur for many other traditional tests as the dimension increases with the sample size. More importantly, as in our paper, investigating these phase transition phenomena of classical tests not only contributes to the theoretical development but also motivates us to propose new test procedure or more advanced approximation distributional theory which are suitable under the high dimensional scenario.  
\qed
\end{remark}


The following lemma establishes an upper bound for $\Delta_{q}$.


\begin{lemma}
\label{Lemma_Delta_q}
Assuming that $M_{q} < \infty$, then we have
\begin{align*}
    \Delta_{q} < 2\left(\frac{1}{m} \max_{1\leq i\leq n} P_{ii}\right)^{\delta/2} M_{q}. 
\end{align*}
\end{lemma}

\begin{remark}
\label{Remark_linear_model}
Condition~\eqref{cond_Delta_q} can be viewed as the Lyapunov-type condition for high dimensional Gaussian approximation of $Q_{n}$. It is quite natural and does not impose any explicit restriction on the relation between the dimension $d$ and the sample size $n$ directly. In particular, \eqref{cond_Delta_q} can be dimension free for some commonly used models, namely, \eqref{cond_Delta_q} holds for arbitrary dimension $d \geq 1$ as long as $n \to \infty$. For instance, suppose that $\{V_{i}\}_{i = 1}^{n}$ follow the linear process model
\begin{align}
\label{linearmodel}
    V_{i} = A \xi_{i} \enspace (i = 1, \ldots, n),
\end{align}
where $A$ is a $d\times L$ matrix for some integer $L \geq 1$, $\xi_{i} = (\xi_{i1}, \ldots, \xi_{iL})^{\top}$ and $\{\xi_{i\ell}\}_{i, \ell\in \mathbb{N}}$ are independent zero-mean random variables with uniformly bounded~\emph{q}th moment $\mathbb{E}|\xi_{i\ell}|^{q} \leq C < \infty$. Applying the Burkholder inequality leads to $M_{q} \leq (1 + \delta)^{q} \max_{1\leq \ell\leq L} \|\xi_{i\ell}\|_{q}^{2q}$. Consequently, Lemma~\ref{Lemma_Delta_q} reveals that a sufficient condition for $\Delta_{q} \to 0$ is
\begin{align}
\label{cond_sufficient}
    \frac{1}{m} \max_{1\leq i\leq n} P_{ii} \to 0.
\end{align}
It is worth mentioning that~\eqref{cond_sufficient} depends only on the projection matrix $P$ and does not impose any restriction on the dimension $d$. Moreover, under Assumption~\ref{cond_design}, \eqref{cond_sufficient} is automatically satisfied in view of $\max_{1\leq i\leq n} (P_{ii}/m)^{2} \leq m^{-2} \sum_{i = 1}^{n} P_{ii}^{2} \to 0$. 
\qed
\end{remark}

\subsection{Modified~\emph{U}~type test statistics}
\label{secU}
The dichotomous nature of the asymptotic null distribution makes $Q_{n}$ unsuitable for testing~\eqref{test} in the high dimensional setting. This motivates us to propose a modified~\emph{U}~type test statistic of $Q_{n}$ for which such a dichotomy does not occur. To fix the idea, let $B_{0} \in \bbR^{p\times d}$ denote the coefficient matrix of model~\eqref{model matrix} under the null hypothesis such that $\mathcal{C} B_{0} = 0$ and $Y \overset{H_{0}}{=} X B_{0} + V$. Motivated by Theorem~\ref{GArho}, a natural candidate of the test statistic $Q_{n}$ would be
\begin{align}
\label{eq_tilde_Qn}
    Q_{n, 0} = Q_{n} - \sum_{k = 1}^{n} P_{kk} (Y_{k} - B_{0}^{\top} X_{k})^{\top} (Y_{k} - B_{0}^{\top} X_{k}),
\end{align}
which coincides with $Q_{n}^{\star}$ in~\eqref{eq_decomposition_Q_n} under the null hypothesis. However, $B_{0}$ is unknown in practice and hence $Q_{n, 0}$ is infeasible. The primary goal of this section is to propose a consistent empirical approximation $U_{n}$ for $Q_{n, 0}$. In particular, motivated by the discussions in Section~\ref{secV}, the modified test statistic $U_{n}$ should satisfy that 
\begin{align*}
    U_{n} \overset{H_{0}}{=} \sum_{i = 1}^{n} \sum_{j \neq i} K_{ij} V_{i}^{\top} V_{j} \enspace \mathrm{and} \enspace \frac{U_{n} - Q_{n, 0}}{\surd{\mathrm{var}(Q_{n, 0})}} \overset{H_{0}}{=} o_{\bbP}(1),
\end{align*}
for some symmetric matrix $K = (K_{ij})_{n \times n}$. The latter ensures that $U_{n}$ is asymptotically equivalent to $Q_{n, 0}$ in~\eqref{eq_tilde_Qn}. Towards this end, let $\hat{B}_{0}$ be the least square estimator of $B$ under the constraint $\mathcal{C}B = 0$. Then $Y - X\hat{B}_{0} = (I_{n} - P_{0})Y$, where $P_{0} = X(X^{\top} X)^{-1} X^{\top} - P$ is the projection matrix of model~\eqref{model matrix} under the null hypothesis. In view of~\eqref{eq_tilde_Qn}, the modified~\emph{U}~type test statistic is then defined by
\begin{align}
\label{eq:M071055}
    U_{n} &= Q_{n} - \sum_{k = 1}^{n} \theta_{k} (Y_{k} - \hat{B}_{0}^{\top} X_{k})^{\top} (Y_{k} - \hat{B}_{0}^{\top} X_{k})\cr 
    &\overset{H_{0}}{=} \sum_{i = 1}^{n} \left(P_{ii} - \sum_{k = 1}^{n} \theta_{k} \bar{P}_{ik, 0}^{2}\right) V_{i}^{\top} V_{i} + \sum_{i = 1}^{n} \sum_{j \neq i} \left(P_{ij} - \sum_{k = 1}^{n} \theta_{k} \bar{P}_{ik, 0}\bar{P}_{jk, 0}\right) V_{i}^{\top} V_{j}\cr
    &= \sum_{i = 1}^{n} \sum_{j \neq i} \left(P_{ij} - \sum_{k = 1}^{n} \theta_{k} \bar{P}_{ik, 0}\bar{P}_{jk, 0}\right) V_{i}^{\top} V_{j},
\end{align}
where $\bar{P}_{0} = I_{n} - P_{0} = (\bar{P}_{ij, 0})_{n \times n}$ and the last equality follows by taking $\theta_{1}, \ldots, \theta_{n}$ to be the solutions of the following linear equations 
\begin{align}
\label{eq_solution_theta}
    \sum_{k = 1}^{n} \bar{P}_{ik, 0}^{2} \theta_{k} = P_{ii} \enspace (i = 1, \ldots, n). 
\end{align}
It is worth mentioning that typically $\theta_{k}$ in~\eqref{eq:M071055} are not $P_{k k}$, as one would naturally like to use in view of~\eqref{eq_tilde_Qn}. We can view~\eqref{eq_solution_theta} as a detailed balanced condition as it removes the diagonals in~\eqref{eq:M071055}. Denote $\theta = (\theta_{1}, \ldots, \theta_{n})^{\top}$ and rewrite~\eqref{eq_solution_theta} in the more compact matrix form
\begin{align}
\label{eq_solution_theta_matrix}
    (\bar{P}_{0} \circ \bar{P}_{0}) \theta = (P_{11}, \ldots, P_{nn})^{\top}.
\end{align}
Let $P_{\theta} = P - \bar{P}_{0} D_{\theta} \bar{P}_{0} = (P_{ij, \theta})_{n \times n}$, where $D_{\theta} = \mathrm{diag}(\theta_{1}, \ldots, \theta_{n})$ is a diagonal matrix. Then $P_{ii, \theta} = 0$ for all $i = 1, \ldots, n$ in view of~\eqref{eq_solution_theta_matrix} and 
\begin{align*}
    U_{n} \overset{H_{0}}{=} \mathrm{tr} (V^{\top}P_{\theta} V) = \sum_{i = 1}^{n} \sum_{j \neq i} P_{ij, \theta} V_{i}^{\top} V_{j}.
\end{align*}
Before proceeding, we first introduce a sufficient condition such that $U_{n}$ exists and is well defined.

\begin{lemma}
\label{tildep}
Assume that there exists a positive constant $\varpi_{0} < 1/2$ such that
\begin{align}
\label{hii0bound}
    \max_{1\leq i\leq n} P_{ii, 0} \leq \varpi_{0}.
\end{align}
Then the matrix $\bar{P}_{0} \circ \bar{P}_{0}$ is strictly diagonally dominant and $|P_{\theta}|_{\mathbb{F}}^{2} = m - \sum_{i = 1}^{n} \theta_{i} P_{ii}$. Moreover, if $\max_{1\leq i\leq n} P_{ii} \leq \varpi_{1} \zeta$ for some positive constant $\varpi_{1} < 1/2$, where $\zeta = (1 - 2\varpi_{0})(1 - \varpi_{0})$, then we have $\max_{1\leq i\leq n} |\theta_{i}| \leq \varpi_{1} < 1/2$. 
\end{lemma}

\begin{remark}
Condition~\eqref{hii0bound} ensures the matrix $\bar{P}_{0} \circ \bar{P}_{0}$ is invertible. Consequently the solution $\theta$ of~\eqref{eq_solution_theta_matrix} exists and is unique. It is worth noting that $\theta$ is independent of the dimension $d$ and only depends on the projection matrices $P$ and $P_{0}$. Moreover, as shown in the proof of Lemma~\ref{tildep},  
\begin{align*}
    \sum_{i = 1}^{n} \theta_{i} P_{ii} \leq \frac{1}{\zeta} \sum_{i = 1}^{n} P_{ii}^{2} \enspace \mathrm{and} \enspace \max_{1\leq i\leq n} |\theta_{i}| \leq \frac{1}{\zeta} \max_{1\leq i\leq n} P_{ii}, 
\end{align*}
which are essential to upper bound the quantity $\Delta_{q, \theta}$ in Lemma~\ref{Lemma_Delta_diamond} below. Consequently, under Assumption~\ref{cond_design}, suppose $\sum_{i = 1}^{n} P_{ii}^{2} \leq m \zeta /2$ for sufficiently large $n$, we obtain  
\begin{align*}
    \var(U_{n}) = 2 |P_{\theta}|_{\mathbb{F}}^{2} \varsigma^{2} = 2\left(m - \sum_{i = 1}^{n} \theta_{i} P_{ii}\right) \varsigma^{2} > m \varsigma^{2},
\end{align*}
which ensures the proposed test statistic $U_{n}$ is non-degenerate and well defined. 
\qed
\end{remark}

\begin{remark}
\label{Remark_cond_balance}
Since $col(X(X^{\top} X)^{-1}\mathcal{C}^{\top}) \subset col(X)$, where $col(\cdot)$ denotes the column space, $P_{0} = X(X^{\top} X)^{-1} X^{\top} - P$ defined above is also a projection matrix. Hence $\max\{P_{ii}, P_{ii, 0}\} \leq X_{i}^{\top}(X^{\top} X)^{-1} X_{i}$ uniformly for $i \in \{1, \ldots, n\}$ and a sufficient condition for Lemma~\ref{tildep} would be
\begin{align}
\label{eq_max_h_ii}
    \max_{1\leq i\leq n} X_{i}^{\top} (X^{\top} X)^{-1} X_{i} \leq \min\{\varpi_{0}, (1 - 2\varpi_{0})(1 - \varpi_{0})\varpi_{1}\}, 
\end{align}
which is fairly mild on the design matrix $X$. More specifically, it is commonly assumed~\citep{Huber1973, Portnoy1985, Wu1986, Shao1987, Shao1988, Mammen1989, Navidi1989, Lahiri1992} for the linear regression model that $\max_{1\leq i\leq n} X_{i}^{\top}(X^{\top} X)^{-1} X_{i} \to 0$, which ensures a kind of ``robustness of design''~\citep{Huber1973}. It also implies Assumption~\ref{cond_design} in view of Remark~\ref{Remark_P_matrix} and can be viewed as a imbalance measure of model~\eqref{model matrix}~\citep{Shao1987}. 
\qed
\end{remark}

\begin{example}
\label{Example_cond_Gaussian}
Suppose $X_{1}, \ldots, X_{n}$ are independent Gaussian random vectors $N(0, \Gamma)$, where the covariance matrix $\Gamma\in \mathbb{R}^{p\times p}$ has minimal eigenvalue $\lambda_{\min}(\Gamma)>0$. Then, with probability at least $1 - 2\exp(-n/2) - n^{-1}$, we have 
\begin{align}
\label{eq_max_Projection_matrix}
    \max_{1\leq i\leq n} X_{i}^{\top}(X^{\top} X)^{-1} X_{i} \leq \frac{9p + 18\sqrt{2p \log n} + 36 \log n}{n}. 
\end{align}
Consequently, condition~\eqref{eq_max_h_ii} holds with high probability as long as $p/n$ is sufficiently small. 
\qed
\end{example}

\begin{proposition}
\label{Proposition_test}
Under the conditions of Lemma~\ref{tildep}, we have $\bbE (U_{n}) \geq 0$. In particular, 
\begin{align*}
    \bbE(U_{n}) = 0 \enspace \mathrm{if\ and\ only\ if} \enspace \mathcal{C}B = 0. 
\end{align*}
\end{proposition}

\subsection{Asymptotic distribution of the modified test statistics}
The primary goal of this section is to establish a Gaussian approximation for the modified test statistic $U_{n}$. Following~\eqref{G0}, the Gaussian analogue of $U_{n}$ is defined by 
\begin{align*}
    \mathcal{G}_{n} = \tr(Z^{\top} P_{\theta} Z) = \sum_{i = 1}^{n} \sum_{j \neq i} P_{ij, \theta} Z_{i}^{\top} Z_{j}.
\end{align*}
The following theorem establishes a non-asymptotic upper bound of the Kolmogorov distance between the distribution functions of $U_{n}$ and its Gaussian analogue $\mathcal{G}_{n}$. Compared with Theorem~\ref{GArho}, it reveals that the modification of the test statistic $Q_{n}$ in~\eqref{eq:M071055} removes the dichotomous nature of its asymptotic null distribution.

\begin{theorem}
\label{GAtheorem}
Let $q = 2 + \delta$, where $0 < \delta \leq 1$. Assume that~\eqref{hii0bound} holds and that
\begin{align*}
    \Delta_{q, \theta} = \frac{\sum_{i = 1}^{n} \sum_{j \neq i} |P_{ij, \theta}|^{q}}{m^{q/2}} M_{q} + \frac{\sum_{i = 1}^{n} (\sum_{j \neq i} P_{ij, \theta}^{2})^{q/2}}{m^{q/2}} L_{q} \to 0. 
\end{align*}
Then, under Assumptions~\ref{cond_design} and the null hypothesis, we have 
\begin{align*}
    \rho(U_{n}, \mathcal{G}_{n}) \leq C_{q} \Delta_{q, \theta}^{1/(2q + 1)} + C \left(\frac{1}{m} \sum_{i = 1}^{n} P_{ii}^{2}\right)^{1/5} \to 0. 
\end{align*}
\end{theorem}

Similar to Lemma~\ref{Lemma_Delta_q}, we establish a similar upper bound for $\Delta_{q, \theta}$ in the following lemma.

\begin{lemma}
\label{Lemma_Delta_diamond}
Under condition~\eqref{hii0bound}, we have 
\begin{align*}
    \Delta_{q, \theta} \lesssim \left(\frac{1}{m} \max_{1\leq i\leq n} P_{ii}\right)^{\delta/2} M_{q}. 
\end{align*}
\end{lemma}

For $\alpha \in (0, 1)$, Proposition~\ref{Proposition_test} and Theorem~\ref{GAtheorem} motivate an $\alpha$ level test for~\eqref{test} as follows,
\begin{align}
\label{test_Phi_theta}
    \Phi_{\theta} = \mathbb{I}\left\{\frac{U_{n}}{\varsigma |P_{\theta}|_{\mathbb{F}}\surd{2}} > c_{1 - \alpha}\right\},
\end{align}
where $c_{1 - \alpha}$ is the $(1 - \alpha)$th quantile of the standardized $\mathcal{G}_{n}/\surd{\mathrm{var}(\mathcal{G}_{n})}$.

\begin{remark}
It is worth mentioning that the approximating distribution $\mathcal{G}_{n}$ may or may not be asymptotically normal. Let $\lambda_{1}(P_{\theta}), \ldots, \lambda_{n}(P_{\theta})$ denote the eigenvalues of the symmetric matrix $P_{\theta}$. Being a quadratic functional of Gaussian random vectors $\{Z_{i}\}_{i = 1}^{n}$, $\mathcal{G}_{n}$ is distributed as a linear combination of independent chi-squared random variables,  
\begin{align*}
    \mathcal{G}_{n} \overset{\mathcal{D}}{=} \sum_{k = 1}^{d} \sum_{i = 1}^{n} \lambda_{k}(\Sigma) \lambda_{i}(P_{\theta}) \eta_{ik}(1) = \sum_{k = 1}^{d} \sum_{i = 1}^{n} \lambda_{k}(\Sigma) \lambda_{i}(P_{\theta}) \{\eta_{ik}(1) - 1\}, 
\end{align*}
where $\{\eta_{ik}(1)\}_{i, k \in \mathbb{N}}$ are independent $\chi_{1}^{2}$ random variables and the last equality follows from the fact that $\sum_{i = 1}^{n} \lambda_{i}(P_{\theta}) = \sum_{i = 1}^{n} P_{ii, \theta} = 0$. More specifically, the Lindeberg-Feller central limit theorem and Lemma~\ref{tildep} imply that $\mathcal{G}_{n}/\surd{\var(\mathcal{G}_{n})} \Rightarrow N(0, 1)$ if and only if 
\begin{align}
\label{condclt}
    \frac{\lambda_{1}(\Sigma)}{\varsigma\surd{m}} \to 0.
\end{align}
Consequently, $c_{1 - \alpha}$ in~\eqref{test_Phi_theta} is asymptotically equal to the standard normal quantiles whenever~\eqref{condclt} holds.

When $m \to \infty$, condition~\eqref{condclt} automatically holds for arbitrary dimension $d \geq 1$ as $\lambda_{1}(\Sigma) \leq \varsigma$. Otherwise, \eqref{condclt} is equivalent to $\mathrm{tr}(\Sigma^{4})/\varsigma^{4} \to 0$, which is a common assumption to ensure the asymptotic normality of high dimensional quadratic statistics; see, for example, \citet{Bai1996}, \citet{Chen2010}, \citet{Cai2013}, \citet{yao2018testing} and~\citet{Zhang2018} among others. In particular, it reveals that the asymptotic null distribution of $U_{n}$ can be non-normal if~\eqref{condclt} is violated. For example, let $Y_{1}, \ldots, Y_{n} \in \mathbb{R}^{d}$ be i.i.d.~random vectors with mean vector $\mu_{Y} = \mathbb{E}(Y_{1})$ and consider testing whether $\mu_{Y} = 0$. Assume that $\Sigma = \cov(Y_{1}) = (\Sigma_{jk})_{d \times d}$ has entries $\Sigma_{jk} = \vartheta + (1 - \vartheta)\mathbb{I}\{j = k\}$ for some constant $\vartheta \in (0, 1)$. Then $\lambda_{1}(\Sigma)/(\varsigma\surd{m}) \to 1$ and it follows from Theorem~\ref{GAtheorem} that
\begin{align*}
    \frac{U_{n}}{\surd\mathrm{var}(U_{n})} = \frac{\sum_{i = 1}^{n} \sum_{j \neq i} Y_{i}^{\top} Y_{j}}{\varsigma\surd{\{2n(n - 1)}\}} \Rightarrow \frac{\chi_{1}^{2} - 1}{\surd{2}}.
\end{align*}
The simulation study in Section~\ref{secsimulation} shows that our Gaussian multiplier bootstrap approach have a satisfactory performance regardless of whether $U_{n}$ is asymptotically normal or not.
\qed
\end{remark}

\section{Applications}
As mentioned in the introduction, our paradigm~\eqref{test} is fairly general and it can be applied to many commonly studied hypothesis testing problems. In this section, we consider two specific examples to illustrate the usefulness of the proposed~\emph{U}~type test statistic and the corresponding asymptotic distribution theory.

\subsection{High dimensional one-way MANOVA}
\label{Section_one_way_MANOVA}
Let $\{\mathcal{Y}_{ij}\}_{j = 1}^{n_{i}}$, $i = 1, \ldots, K$, be $K \geq 2$ independent samples following the model 
\begin{align*}
    \mathcal{Y}_{ij} = \mu_{i} + \mathcal{V}_{ij} \enspace (j = 1, \ldots, n_{i};\ i = 1, \ldots, K),
\end{align*}
where $\mu_{1}, \ldots, \mu_{K} \in \mathbb{R}^{d}$ are unknown mean vectors of interest, $\{\mathcal{V}_{ij}\}_{j \in \mathbb{N}}$ are i.i.d.~\emph{d}-dimensional random vectors with $\bbE (\mathcal{V}_{i1}) = 0$ and $\cov(\mathcal{V}_{i1}) = \Sigma$. We are interested in testing the equality of the $K$ mean vectors, namely, testing the hypotheses
\begin{align*}
    H_{0} : \mu_{1} = \ldots = \mu_{K} \enspace \mathrm{versus} \enspace H_{1} : \mu_{i} \neq \mu_{l} \enspace \mathrm{for \ some}\enspace 1\leq i \neq l\leq K.
\end{align*}
Following the construction of~\eqref{eq:M071055}, we propose the~\emph{U}~type test statistic 
\begin{align}
\label{test_statistic_MANOVA}
    U_{n K} = \sum_{i = 1}^{K} P_{ii, K} \sum_{j = 1}^{n_{i}} \sum_{k \neq j} \mathcal{Y}_{ij}^{\top} \mathcal{Y}_{ik} + \sum_{i = 1}^{K} \sum_{l \neq i} P_{il, K} \sum_{j = 1}^{n_{i}}\sum_{k = 1}^{n_{l}} \mathcal{Y}_{ij}^{\top} \mathcal{Y}_{lk},
\end{align}
where $n = \sum_{i = 1}^{K} n_{i}$ is the total sample size,
\begin{align*}
    P_{ii, K} = \frac{1}{n - 2} \left(\frac{n}{n_{i}} - \frac{n + K - 2}{n - 1}\right) \enspace \mathrm{and} \enspace P_{il, K} = \frac{1}{n - 2} \left(\frac{1}{n_{i}} + \frac{1}{n_{l}} - \frac{n + K - 2}{n - 1}\right).
\end{align*}
In the context of two sample test for mean vectors where $K = 2$, $U_{nK}$ in~\eqref{test_statistic_MANOVA} reduces to
\begin{align*}
    U_{n K} = \frac{\sum_{i = 1}^{n_{1}} \sum_{j \neq i} \sum_{k = 1}^{n_{2}} \sum_{l \neq k} (\mathcal{Y}_{1i} - \mathcal{Y}_{2k})^{\top}(\mathcal{Y}_{1j} - \mathcal{Y}_{2l})}{(n - 1)(n - 2) n_{1} n_{2}/n},
\end{align*}
which coincides with the commonly used~\emph{U}~type test statistic~\citep{Chen2010}.

For each $i \in \{1, \ldots, K\}$, let $\{\mathcal{Z}_{ij}\}_{j \in \mathbb{N}}$ be i.i.d.~centered Gaussian random vectors with covariance matrix $\cov(\mathcal{Z}_{ij}) = \Sigma$. Following~\eqref{G0}, the Gaussian analogue of $U_{n K}$ is defined by 
\begin{align*}
    \mathcal{G}_{n K} = \sum_{i = 1}^{K} P_{ii, K} \sum_{j = 1}^{n_{i}} \sum_{k \neq j} \mathcal{Z}_{ij}^{\top} \mathcal{Z}_{ik} + \sum_{i = 1}^{K} \sum_{l \neq i} P_{il, K} \sum_{j = 1}^{n_{i}}\sum_{k = 1}^{n_{l}} \mathcal{Z}_{ij}^{\top} \mathcal{Z}_{lk}.
\end{align*}
Let $n_{\min} = \min_{1\leq l\leq K} n_{l}$. Since $\max_{1\leq i\leq n} P_{ii} \leq n_{\min}^{-1}$, Assumption~\ref{cond_design} holds as long as $n_{\min} \to \infty$. The following proposition establishes a non-asymptotic upper bound on the Kolmogorov distance between the distribution functions of $U_{nK}$ and $\mathcal{G}_{nK}$.

\begin{proposition}
Let $q = 2 + \delta$ for some $0 < \delta \leq 1$. Assume that $n_{\min} \to \infty$ and
\begin{align*}
    \tilde{M}_{q} = \max_{1\leq l, l'\leq K} \mathbb{E} \left|\frac{\mathcal{V}_{l1}^{\top}\mathcal{V}_{l'2}}{\varsigma}\right|^{q} < \infty, \enspace \mathrm{where} \enspace \varsigma = |\Sigma|_{\mathbb{F}}.
\end{align*}
Then, under the null hypothesis, we have 
\begin{align*}
    \rho(U_{n K}, \mathcal{G}_{n K}) \leq C_{q} \left(\tilde{M}_{q} n_{\min}^{-\delta/2}\right)^{1/(2q + 1)}\rightarrow 0. 
\end{align*}
\end{proposition}

\begin{remark}
\label{Remark_one_way_MANOVA}
It is worth mentioning that both the dimension $d$ and the number of groups $K$ can grow with the total sample size $n$. In particular, as discussed in Remark~\ref{Remark_linear_model}, if all the $K$ samples follow the linear process model in~\eqref{linearmodel}, $\rho(U_{nK}, \mathcal{G}_{nK}) \to 0$ as long as $n_{\min} \to \infty$. 
\qed
\end{remark}

\subsection{High dimensional nonparametric one-way MANOVA}
For each $i \in \{1, \ldots, K\}$, let $F_{i}$ denote the distribution function of $\mathcal{Y}_{i1}$. We consider testing whether these $K$ independent samples are equally distributed, namely, testing the hypotheses 
\begin{align}
\label{eq_nonparametric_MANOVA}
    H_{0} : F_{1} = \ldots = F_{K} \enspace \mathrm{versus} \enspace H_{1} : F_{i} \neq F_{l} \enspace \mathrm{for\ some} \enspace 1\leq i\neq l \leq K.
\end{align}
Being fundamental and important in statistical inference, \eqref{eq_nonparametric_MANOVA} has been extensively studied; see, for example, \citet{Kruskal1952}, \citet{Akritas1994}, \citet{Brunner2001}, \citet{Rizzo2010} and~\citet{Thas2010} among many others. However, all the aforementioned works mainly focus on the traditional low dimensional scenario and testing~\eqref{eq_nonparametric_MANOVA} for high dimensional random vectors has been much less studied. In this section, we propose a new~\emph{U} type test statistic for~\eqref{eq_nonparametric_MANOVA} following the intuition of~\eqref{eq:M071055} and establish the corresponding distributional theory. In particular, our asymptotic framework is fairly general and allows both the dimension $d$ and the number of groups $K$ to grow with $n$.


To begin with, for each $i \in \{1, \ldots, K\}$, let $\phi_{i}(t) = \bbE \{\exp(\imath t^{\top} \mathcal{Y}_{ij})\}$ denote the characteristic function of $\mathcal{Y}_{ij}$, where $\imath$ stands for the imaginary unit. Then it is equivalent to test the hypotheses
\begin{align}
\label{eq_Hypothesis_phi}
    H_{0} : \phi_{1} = \ldots = \phi_{K} \enspace \mathrm{versus} \enspace H_{1} : \phi_{i} \neq \phi_{l} \enspace \mathrm{for\ some} \enspace 1\leq i\neq l \leq K.  
\end{align}
Denote $\mathcal{Y}_{ij}(t) = \exp(\imath t^{\top}\mathcal{Y}_{ij})$. Similar to~\eqref{test_statistic_MANOVA}, our test statistic for~\eqref{eq_Hypothesis_phi} is defined by 
\begin{align*}
    \tilde{U}_{n K} = \sum_{i = 1}^{K} P_{ii, K} \sum_{j = 1}^{n_{i}} \sum_{k \neq j} \int \mathcal{Y}_{ij}(t) \overline{\mathcal{Y}_{ik}(t)} w(t) dt + \sum_{i = 1}^{K} \sum_{l \neq i} P_{il, K} \sum_{j = 1}^{n_{i}} \sum_{k = 1}^{n_{l}} \int \mathcal{Y}_{ij}(t) \overline{\mathcal{Y}_{lk}(t)} w(t) dt, 
\end{align*}
where $w(t) \geq 0$ is a suitable weight function such that the integrals above are well defined. Discussions of some commonly used weight functions are given in Remark~\ref{Remark_weight_function} below.

Before proceeding, we first define the Gaussian analogue of $\tilde{U}_{nK}$ under the null hypothesis that the $K$ samples are equally distributed. Define the covariance function of $\mathcal{Y}_{11}(t)$ as
\begin{align*}
    \Sigma(t, s) = \mathbb{E}\{\mathcal{Y}_{11}(t) - \phi_{1}(t)\}\overline{\{\mathcal{Y}_{11}(s) - \phi_{1}(s)\}} = \phi_{1}(t - s) - \phi_{1}(t)\phi_{1}(-s) \enspace (t, s \in \mathbb{R}^{d}). 
\end{align*}
Throughout this section, by Mercer's theorem, we assume that the covariance function above admits the following eigendecomposition 
\begin{align*}
    \Sigma(t, s) = \sum_{m = 1}^{\infty} \lambda_{m} \varphi_{m}(t)\overline{\varphi_{m}(s)} \enspace (t, s \in \mathbb{R}^{d}),  
\end{align*}
where $\lambda_{1} \geq \lambda_{2} \geq \ldots \geq 0$ are eigenvalues and $\varphi_{1}, \varphi_{2}, \ldots$, are the corresponding eigenfunctions. We now apply the Karhunen–Loève theorem. Let $\{Z_{ijk}\}_{i, j, k \in \mathbb{N}}$ be independent~standard normal random variables and define Gaussian processes  
\begin{align*}
    \mathcal{Z}_{ij}(t) = \sum_{m = 1}^{\infty} \surd{\lambda_{m}} Z_{ijm} \varphi_{m}(t) \enspace (t \in \mathbb{R}^{d}). 
\end{align*}
Then, following~\eqref{G0}, the Gaussian analogue of $\tilde{U}_{nK}$ is defined by 
\begin{align*}
    \tilde{\mathcal{G}}_{n K} = \sum_{i = 1}^{K} P_{ii, K} \sum_{j = 1}^{n_{i}} \sum_{k \neq j} \int \mathcal{Z}_{ij}(t) \overline{\mathcal{Z}_{ik}(t)} w(t) dt + \sum_{i = 1}^{K} \sum_{l \neq i} P_{il, K} \sum_{j = 1}^{n_{i}} \sum_{k = 1}^{n_{l}} \int \mathcal{Z}_{ij}(t) \overline{\mathcal{Z}_{lk}(t)} w(t) dt. 
\end{align*}

\begin{proposition}
Let $q = 2 + \delta$ for some $0 < \delta \leq 1$. Assume that $n_{\min} \to \infty$ and  
\begin{align*}
    \tilde{\mathcal{M}}_{q} = \bbE \left|\frac{\int_{\mathbb{R}^{d}} \mathbb{E}\{\mathcal{Y}_{11}(t)\} \overline{\mathbb{E}_{0}\{\mathcal{Y}_{12}(t)\}} w(t) dt}{\mathcal{F}}\right|^{q} < \infty, \enspace \mathrm{where} \enspace \mathcal{F}^{2} = \sum_{m = 1}^{\infty} \lambda_{m}^{2}. 
\end{align*}
Then, under the null hypothesis that these $K$ independent samples are equally distributed, we have 
\begin{align*}
    \rho(\tilde{U}_{n K}, \tilde{\mathcal{G}}_{n K}) \leq C_{q} \left(\tilde{\mathcal{M}}_{q} n_{\min}^{-\delta/2}\right)^{1/(2q + 1)}\rightarrow 0. 
\end{align*}
\end{proposition}

\begin{remark}
\label{Remark_weight_function}
It is worth mentioning that the proposed test statistic $\tilde{U}_{n K}$ contains high dimensional integral over $t \in \mathbb{R}^{d}$, which can be computational intractable in practice. To make $\tilde{U}_{n K}$ well defined and facilitate the computation, we shall choose suitable weight function $w(t)$ such that $\tilde{U}_{n K}$ has a simple closed-form expression. In the literature, various kinds of weight functions have been proposed such as the Gaussian kernel function~\citep{Gretton2012}, the Laplace kernel function~\citep{Gretton2012} and the energy kernel function~\citep{Szekely2007, Rizzo2010}. For instance, let $w(t)$ denote the density function of the random vector $\mathcal{X} \kappa/\surd{\eta}$ for some $\kappa > 0$, where $\mathcal{X} \sim N(0, I_{d})$ and $\eta \sim \chi_{1}^{2}$ are independent (equivalently $\mathcal{X} \kappa/\surd{\eta}$ is a Cauchy random variable with location parameter 0 and scale parameter $\kappa$). Then it is straightforward to verify that
\begin{align*}
    \int \mathcal{Y}_{ij}(t) \overline{\mathcal{Y}_{lk}(t)} w(t) dt = \int \cos\{t^{\top}(Y_{ij} - Y_{lk})\} w(t) dt = \exp(-\kappa |Y_{ij} - Y_{lk}|), 
\end{align*}
which is the same as the Laplace kernel function with $1/\kappa$ being its bandwidth, where $|\cdot|$ stands for the Euclidean distance. A more general result can be derived using Bochner's Theorem, see e.g., Theorem 3.1 of \cite{Gretton2009}. Consequently, the proposed test statistic $\tilde{U}_{n K}$ reduces to 
\begin{align*}
    \tilde{U}_{n K} = \sum_{i = 1}^{K} P_{ii, K} \sum_{j = 1}^{N_{i}} \sum_{k \neq j} \exp(-\kappa|Y_{ij} - Y_{ik}|) + \sum_{i = 1}^{K} \sum_{l \neq i} P_{il, K} \sum_{j = 1}^{N_{i}} \sum_{k = 1}^{N_{l}} \exp(-\kappa|Y_{ij} - Y_{lk}|),  
\end{align*}
which is fairly convenient to compute in practice. Moreover, suitable choice of the weight function $w(t)$ also facilitate the analysis of the quantities $\mathcal{M}_{q}$ and $\mathcal{F}$.
\qed   
\end{remark}

\section{Practical implementation}
\label{secsigma}
In this section, we propose an unbiased estimator for $\varsigma^{2}$, which is ratio-consistent under fairly mild moment conditions. To begin with, since $\mathbb{E}(V_{i}^{\top} V_{j})^{2} = \varsigma^{2}$ for any $i\neq j$, a natural unbiased~\emph{U} type estimator for $\varsigma^{2}$ based on $\{V_{i}\}_{i = 1}^{n}$ would be 
\begin{align}
\label{eq_oracle_estimator_spectral}
    \hat{\varsigma}_{o}^{2} = \frac{1}{n(n - 1)} \sum_{i = 1}^{n} \sum_{j \neq i} (V_{i}^{\top} V_{j})^{2}. 
\end{align}
Let $\bar{P}_{1} = I_{n} - X(X^{\top} X)^{-1} X^{\top} = (P_{ij, 1})_{n \times n}$ and $\hat{V} = \bar{P}_{1} Y = (\hat{V}_{1}, \ldots, \hat{V}_{n})^{\top}$. It is worth noting that directly substituting the residual vectors $\{\hat{V}_{i}\}_{i = 1}^{n}$ into~\eqref{eq_oracle_estimator_spectral} yields a feasible but generally biased estimator for $\varsigma^{2}$. More specifically, for any $i \neq j$, 
\begin{align*}
    \mathbb{E} (\hat{V}_{i}^{\top}\hat{V}_{j})^{2} = (\bar{P}_{ii, 1} \bar{P}_{jj, 1} + \bar{P}_{ij, 1}^{2}) \varsigma^{2} + \bar{P}_{ij, 1}^{2} \mathbb{E}(V_{1}^{\top} V_{1})(V_{2}^{\top} V_{2}) + \sum_{k = 1}^{n} (\bar{P}_{ik, 1}\bar{P}_{jk, 1})^{2} \left\{\|\mathbb{E}_{0}(V_{1}^{\top} V_{1})\|_{2}^{2} - 2 \varsigma^{2}\right\},
\end{align*}
which reveals that $(\hat{V}_{i}^{\top} \hat{V}_{j})^{2}$ is no longer unbiased of $\varsigma^{2}$ even after proper scaling. This motivates us to propose a new unbiased estimator for $\varsigma^{2}$ via data-splitting, which excludes the bias terms $(V_i^\top V_i)^2$ and $(V_i^\top V_i)(V_j^\top V_j)$. Without loss of generality, we assume that the sample size $n$ is even in what follows. 
\begin{enumerate}
    \item Randomly split $\{1, \ldots, n\}$ into two halves $\mathcal{A}$ and $\mathcal{A}^{c}$. Denote $\mathcal{M}_{\mathcal{A}} = \{(X_{i}, Y_{i}), i \in \mathcal{A}\}$ and $\mathcal{M}_{\mathcal{A}^{c}} = \{(X_{i}, Y_{i}), i \in \mathcal{A}^{c}\}$. 
    \item For both $\mathcal{M}_{A}$ and $\mathcal{M}_{\mathcal{A}^{c}}$, fit model~\eqref{model1} with the least squares estimates and compute 
    \begin{align*}
        \hat{\Sigma}_{\mathcal{A}} = \frac{1}{n/2 - p} \hat{V}_{\mathcal{A}}^{\top} \hat{V}_{\mathcal{A}} \enspace \mathrm{and} \enspace \hat{\Sigma}_{\mathcal{A}^{c}} = \frac{1}{n/2 - p} \hat{V}_{\mathcal{A}^{c}}^{\top}\hat{V}_{\mathcal{A}^{c}}, 
    \end{align*}
    where $\hat{V}_{\mathcal{A}}$ and $\hat{V}_{\mathcal{A}^{c}}$ are the residual matrices of $\mathcal{M}_{\mathcal{A}}$ and $\mathcal{M}_{\mathcal{A}^{c}}$, respectively.  
    \item Compute the estimator $\hat{\varsigma}_{\mathcal{A}}^{2} = \mathrm{tr}(\hat{\Sigma}_{\mathcal{A}}\hat{\Sigma}_{\mathcal{A}^{c}})$.  
\end{enumerate}
Since $\hat{\Sigma}_\mathcal{A}$ and $\hat{\Sigma}_{\mathcal{A}^{c}}$ are independent and both of them are unbiased estimators of $\Sigma$, $\hat{\varsigma}_\mathcal{A}^{2}$ is unbiased for $\varsigma^{2}$ as $\bbE (\hat{\varsigma}_\mathcal{A}^{2}) = \tr\{\bbE (\hat{\Sigma}_\mathcal{A}) \bbE (\hat{\Sigma}_{\mathcal{A}^c})\} = \tr(\Sigma^2) = \varsigma^{2}$.

\begin{theorem}
\label{splitsigma}
Assume that $p/n < \varpi_{2}$ for some positive constant $\varpi_{2} < 1/2$ and that the least squares estimates are well defined for both $\mathcal{M}_\mathcal{A}$ and $\mathcal{M}_{\mathcal{A}^c}$. Then we have 
\begin{align*}
    \mathbb{E} \left|\frac{\hat{\varsigma}_\mathcal{A}}{\varsigma} - 1\right|^{2} \lesssim \frac{M_{4}}{n^{2}} + \frac{p\times \tr(\Sigma^{4})}{n^{2} \varsigma^{4}} + \frac{\|\mathbb{E}_{0}(V_{1}^{\top}\Sigma V_{1})\|_{2}^{2}}{n \varsigma^{4}}.
\end{align*}
\end{theorem}

\begin{remark}
The proof of Theorem~\ref{splitsigma} is given in Section $7.2$, where a more general upper bound on $\bbE |\hat{\varsigma}_{\mathcal{A}}/\varsigma - 1|^{\tau}$ is established for $1< \tau \leq 2$. Theorem~\ref{splitsigma} reveals that $\hat{\varsigma}_\mathcal{A}$ is ratio consistent under mild moment conditions. Suppose now $\{V_{i}\}_{i\in \mathbb{N}}$ follow the linear process model~\eqref{linearmodel} with $\max_{1\leq \ell \leq L} \mathbb{E}|\xi_{i\ell}|^{4} \leq C <\infty$. Then $M_{4}$ is bounded and $\|\mathbb{E}_{0} (V_{1}^{\top} \Sigma V_{1})\|_{2}^{2} \lesssim \tr(\Sigma^{4})$. Consequently, 
\begin{align*}
    \mathbb{E}\left|\frac{\hat{\varsigma}_{\mathcal{A}}}{\varsigma} - 1\right|^{2} \lesssim n^{-2} + \frac{\tr(\Sigma^{4})}{n \varsigma^{4}}. 
\end{align*}
In this case, $\hat{\varsigma}_{\mathcal{A}}$ is ratio consistent for arbitrary dimension $d \geq 1$ as long as $n \to \infty$. 
\qed
\end{remark}

\begin{remark}
There are totally ${n \choose n/2}$ different ways of splitting $\{1, \ldots, n\}$ into two halves. To reduce the influence of randomness of an arbitrary splitting, we can repeat the procedure independently for multiple times and then take the average of the resulting estimators. We refer to~\citet{Fan2012} for more discussions about data-splitting and repeated data-splitting.
\qed
\end{remark}

\begin{remark}
Let $\hat{\Sigma} = (n - p)^{-1}\hat{V}^\top \hat{V}$. Observe that $\bbE (\hat{V}_{i}^{\top} \hat{V}_{j}) = \bar{P}_{ij, 1} \tr(\Sigma)$. We can estimate $\varsigma^{2}$ via 
\begin{align*}
    \hat{\varsigma}_{S}^{2} = \frac{\sum_{i, j = 1}^{n} |\hat{V}_{i}^{\top}\hat{V}_{j} - \bar{P}_{ij, 1} \tr(\hat{\Sigma})|^{2}}{(n - p + 2)(n - p - 1)} = \frac{(n - p)^{2}}{(n - p + 2)(n - p - 1)}\left[|\hat{\Sigma}|_{\bbF}^{2} - \frac{\{\tr(\hat{\Sigma})\}^{2}}{n - p}\right], 
\end{align*}
which is same as the estimator proposed in~\citet{Srivastava2006}, where $\{V_{i}\}_{i = 1}^{n}$ are assumed to be Gaussian random vectors. See also~\citet{Bai1996}. However, for non-Gaussian $\{V_{i}\}_{i = 1}^{n}$ such that $\|\bbE_{0}(V_{1}^{\top} V_{1})\|_{2}^{2} \neq 2 \varsigma^{2}$, this estimator is generally biased as
\begin{align*}
    \mathbb{E}(\hat{\varsigma}_{S}^{2}) - \varsigma^{2} = \frac{\sum_{i = 1}^{n} \bar{P}_{ii, 1}^{2}}{(n - p)(n - p + 2)} \left\{\|\mathbb{E}_{0}(V_{1}^{\top} V_{1})\|_{2}^{2} - 2\varsigma^{2}\right\}. 
\end{align*}
In particular, the bias of $\hat{\varsigma}_{S}^{2}$ can diverge when $\|\bbE_{0}(V_{1}^{\top} V_{1})\|_{2}^{2}$ is much larger than $\varsigma^{2}$. Below we provide an example that typifies the diverging bias.
\qed
\end{remark}

\begin{figure}
    \centering
    \includegraphics[width=0.75\textwidth]{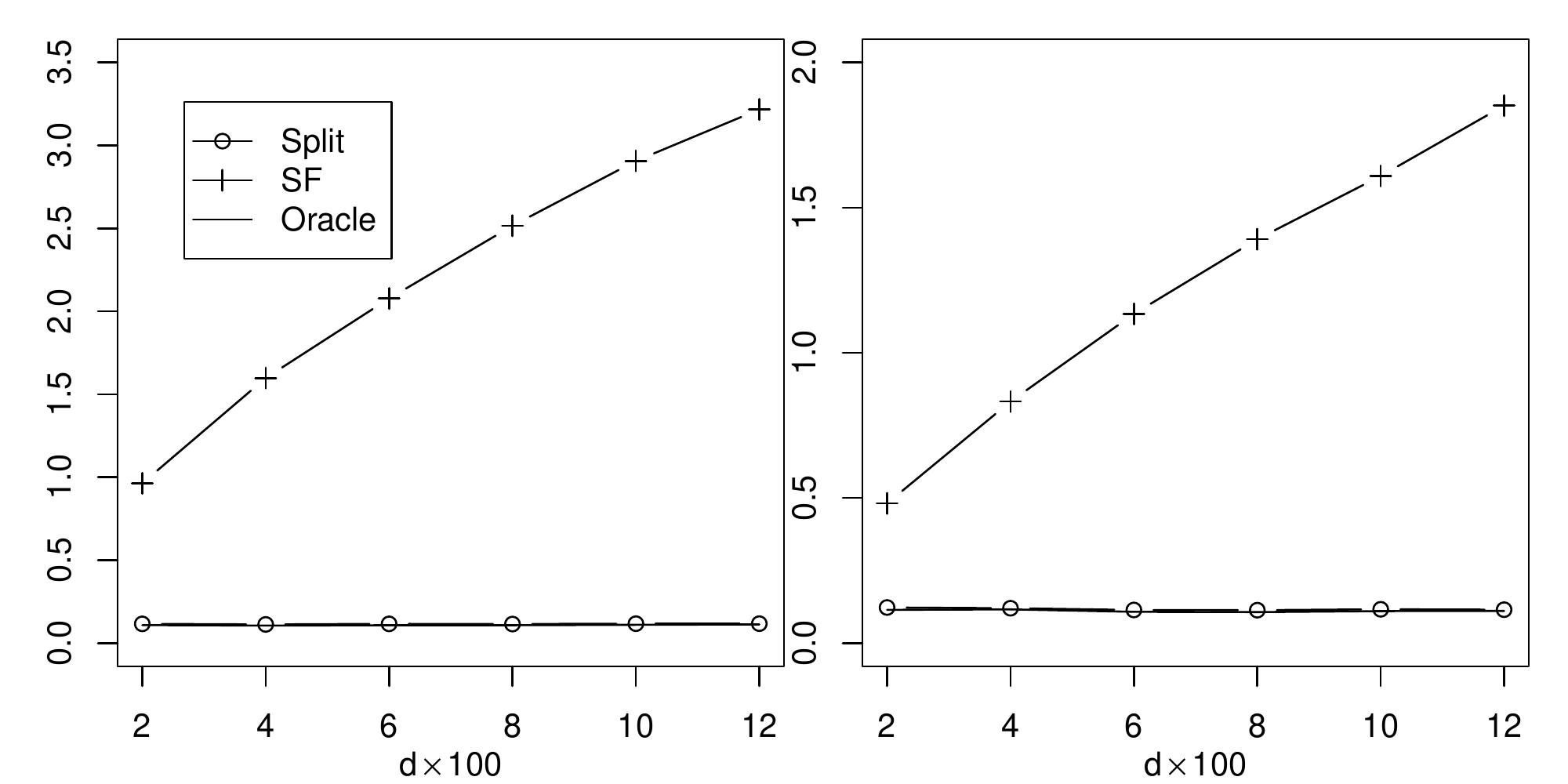}
    \caption{Empirical averages of the values of $|\hat{\varsigma}/\varsigma - 1|$}
    \label{figuresigma}
\end{figure}

\begin{example}
\label{example3}
Let $\{\xi_{i}\}_{i \in \mathbb{N}}$ and $\{\xi_{i}'\}_{i \in \mathbb{N}}$ be two sequences of independent Gaussian random vectors $N(0, \Sigma)$, where $\Sigma = (\Sigma_{ij})_{n \times n}$ has entries $\Sigma_{ij} = \vartheta^{|i - j|}$ for some $\vartheta\in (0, 1)$. Following~\citet{Wang2015}, we draw i.i.d.~innovations $\{V_{i}\}_{i = 1}^{n}$ from a scale mixture of two independent multivariate Gaussian distributions as follows,
\begin{align*}
    V_{i} = \nu_{i} \times \xi_{i} + 3 (1 - \nu_{i}) \times \xi_{i}' \enspace (i = 1, \ldots, n),
\end{align*}
where $\{\nu_{i}\}_{i \in \mathbb{N}}$ are independent Bernoulli random variables with $\pr(\nu_{i} = 1) = 0.9$. A simulation study is given in Section~\ref{secsimulation} by setting $\vartheta = 0.3$ and $0.7$. We report in Figure~\ref{figuresigma} the average values of $|\hat{\varsigma}/\varsigma - 1|$ for $\hat{\varsigma}_{\mathcal{A}}$, $\hat{\varsigma}_{o}$ and $\hat{\varsigma}_{S}$, based on $1000$ replications with the numerical setup $(n, p, m) = (100, 20, 10)$ and $d = 200, 400, 800, 1000, 1200$. For both cases of $\vartheta$, $|\hat{\varsigma}_{\mathcal{A}}/\varsigma - 1|$ and $|\hat{\varsigma}_{o}/\varsigma - 1|$ are very close to $0$, while $|\hat{\varsigma}_{S}/\varsigma - 1|$ is quite large. More precisely, we can derive that $\|\mathbb{E}_{0}(V_{1}^{\top} V_{1})\|_{2}^{2} \approx (18 + d) \varsigma^{2}$.  
\qed
\end{example}


Substituting the ratio-consistent estimator $\hat{\varsigma}_{\mathcal{A}}^{2}$ into $\var(U_{n}) = 2 |P_{\theta}|_{\mathbb{F}}^{2} \varsigma^{2}$ yields $U_{n}/(\hat{\varsigma}_{\mathcal{A}}|P_{\theta}|_{\mathbb{F}}) \Rightarrow N(0, 2)$ under~\eqref{condclt}. Then, for $\alpha \in (0, 1)$, an asymptotic $\alpha$ level test is given by
\begin{align}
\label{testusingclt}
    \Phi_{Z} = \mathbb{I}\left\{\frac{U_{n}}{\hat{\varsigma}_{\mathcal{A}} |P_{\theta}|_{\mathbb{F}}\surd{2}} > z_{1 - \alpha}\right\}, 
\end{align}
where $z_{1 - \alpha}$ is the $(1 - \alpha)$th quantile of the standard normal distribution. 

\section{A simulation study}
\label{secsimulation}
In this section, we conduct a Monte Carlo simulation study to assess the finite sample performance of the proposed tests. In the model~\eqref{model1}, we write $X_i = (1, \mathbf{x}_{i}^\top)^\top \in \mathbb{R}^{p}$ to include an intercept. Here $\mathbf{x}_1, \ldots, \mathbf{x}_n\in\R^{p - 1}$ are i.i.d.~$N(0,I_{p - 1})$ random vectors. Let $m < p$. For all $k \in \{1, \ldots, p - m\}$, all entries of the coefficient vector $B_k$ are i.i.d.~uniform random variables in the interval $(1,2)$. After those $B_k$'s are generated, we keep their values throughout
the simulation. Our goal is to identify the zero $B_k$'s by testing 
\begin{align*}
    H_{0} : B_{p - m + 1} = B_{p - m + 2} = \cdots = B_{p} = 0. 
\end{align*}
In our simulation, we set $(p, m) = (20, 10)$, $n = 100, 200$ and $d = 400, 800, 1200$. We consider two different designs of the innovations $(V_{i})$: the one introduced in Example~\ref{example3} and the one in Example~\ref{example1} below. In both examples, the parameter $\vartheta$ is set to be $0.3$ and $0.7$.
\begin{example}
\label{example1}
Let $\{\xi_{ij}\}_{i, j \in \mathbb{N}}$ be i.i.d.~random variables with $\bbE (\xi_{11}) = 0$ and $\var(\xi_{11}) = 1$. In particular, we consider two cases for $(\xi_{ij})$; they are drawn from the standardized $t_{5}$ distribution and the standardized $\chi_{5}^{2}$ distribution, respectively. For some $\vartheta \in (0, 1)$, we generate
\begin{align*}
    V_{i} = \surd(1 - \vartheta) \times \xi_{i} + \surd\vartheta \times (\xi_{i0}, \xi_{i0}, \ldots, \xi_{i0})^\top, \enspace i \in \mathbb{N}.
\end{align*}
\end{example}

We shall apply a Gaussian multiplier bootstrap approach to implement our proposed test. The procedure is as follows. 
\begin{enumerate}
    \item Compute the residual matrix $\hat V  = (\hat{V}_{1}, \ldots, \hat{V}_{n})^{\top} =  \bar{P}_{1} Y$. Generate i.i.d.~$N(0, 1)$ random variables $\{\omega_{ij}\}_{i, j \in \mathbb{N}}$ and compute the bootstrap residuals $V^{\star} = (V_{1}^{\star}, \ldots, V_{n}^{\star})^{\top}$, where
    \begin{align*}
        V_{i}^{\star} = \frac{1}{\surd(n - p)} \sum_{j = 1}^{n} \omega_{ij} \hat{V}_{i} \enspace (i = 1, \ldots, n).
    \end{align*}
    \item Use $V^{\star}$ to compute $\hat{\varsigma}_{\mathcal{A}}^{\star}$ and the bootstrap test statistic $U_{n}^{\star} = \mathrm{tr}(V^{\star \top} P_{\theta} V^{\star})$. 
    \item Repeat the first two steps independently $\mathcal{B}$ times and collect $U_{n k}^{\star}$ and $\hat{\varsigma}_{\mathcal{A} k}^{\star}$, $k = 1, \ldots, \mathcal{B}$.   
    \item Let $\hat{c}_{1 - \alpha}$ be the $(1 - \alpha)$th quantile of $\{U_{n k}^{\star}/(\hat{\varsigma}_{\mathcal{A} k}^{\star} |P_{\theta}|_{\mathbb{F}}\surd{2})\}_{k = 1, \ldots, \mathcal{B}}$. The our test is 
    \begin{align}
    \label{GMBtest}
        \Phi_{B} = \mathbb{I}\left\{\frac{U_{n}}{\hat{\varsigma}_{\mathcal{A}}|P_{\theta}|_{\mathbb{F}}\surd{2}} > \hat{c}_{1 - \alpha}\right\}, 
    \end{align}
    and we shall reject the null hypothesis whenever $\Phi_{B} = 1$. 
\end{enumerate}
Similar to $\mathcal{G}_{n}$, $U_{n}^{\star}$ is a quadratic functional of i.i.d.~Gaussian random vectors conditional on $\{X, Y\}$ and is distributed as a linear combination of independent chi-squared random variables. To justify the validity of the proposed Gaussian multiplier bootstrap approach, it suffices to bound the distance between the distribution functions of these two quadratic functionals, which can be established by verifying the normalized consistency~\citep{Xu2014} of the corresponding covariance matrix. However, this can be highly non-trivial in the high dimensional setting and is beyond the scope of current paper. Hence we leave it for future work.

In our simulation, we set the bootstrap size $\mathcal{B} = 1000$. As comparison, we also perform the test suggested in~\eqref{testusingclt} based on the central limit theorem and the one proposed in~\citet{Srivastava2013} which we denote by SK. For each test, we report the empirical size based on $2000$ replications as displayed in Table~\ref{sizeexample3} and Table~\ref{sizeexample1}. The results suggest that our proposed test by using the bootstrap procedure provides the best size accuracy in
general as the empirical sizes are close to the nominal level $\alpha$.
\begin{table}[htbp]
\centering
\caption{Empirical sizes for Example~\ref{example3} with $\alpha = 0.05$}
\label{sizeexample3}
\scalebox{0.9}{\begin{tabular}{ccccccccccc}
\hline\hline
&&&\multicolumn{3}{c}{$\theta=0.3$}&&\multicolumn{3}{c}{$\theta=0.7$}\\
\cline{4-6}\cline{8-10}
$n$& $d$& & CLT & GMB & SK  & &CLT & GMB & SK\\
\hline
100 & 400& &0.057& 0.047& 0.041& &0.059& 0.051 & 0.036 \\
& 800& &0.049& 0.045& 0.033& &0.063& 0.056 & 0.026 \\
& 1200& &0.062&0.055& 0.021 & &0.048& 0.045 & 0.028 \\
\hline
 200 & 400& &0.056& 0.052& 0.042& &0.052& 0.047 & 0.037 \\
& 800& &0.052& 0.049& 0.037& &0.053& 0.050 & 0.033 \\
& 1200& &0.045& 0.044 &0.029 & &0.050& 0.046 & 0.035 \\
\hline\hline
\end{tabular}}
\centering
\end{table}

For Example~\ref{example3}, both of the test by CLT and our Gaussian multiplier bootstrap method have better performance than the SK test since the latter is too conservative as $d$ is large. As expected from our theoretical results, normal approximation can work reasonably well in this design.

\begin{table}[htbp]
\centering
\caption{Empirical sizes for Example~\ref{example1} with $\alpha = 0.05$}
\label{sizeexample1}
\scalebox{0.9}{\begin{tabular}{cccccccccccc}
\hline\hline
&&&&\multicolumn{3}{c}{$t_5$}&&\multicolumn{3}{c}{$\chi_5^2$}\\
\cline{5-7}\cline{9-11}
$\theta$ & $n$& $d$& &CLT & GMB & SK & &CLT & GMB & SK\\
\hline
$0.3$ & 100 & 400& & 0.068& 0.058& 0.023& &0.083& 0.065 & 0.036 \\
 &  & 800& &0.082& 0.066& 0.023& &0.074& 0.058 & 0.016 \\
&  & 1200& &0.082& 0.068& 0.015& &0.067& 0.053 & 0.011 \\
 & 200 & 400& &0.073& 0.059& 0.022& &0.067& 0.054 & 0.018 \\
&  & 800& &0.071& 0.057 & 0.012& &0.074& 0.058 & 0.014 \\
&  & 1200& &0.076& 0.059& 0.011& &0.077& 0.058 & 0.011 \\
\hline
$0.7$ & 100 & 400& &0.074& 0.055 & 0.002& &0.082& 0.062 & 0.002 \\
 &  & 800& &0.084& 0.066& 0.001& &0.085& 0.071 & 0.000 \\
&  & 1200& &0.073& 0.057& 0.000& &0.076& 0.062 & 0.001 \\
 & 200 & 400& &0.083& 0.067& 0.001& &0.080& 0.064 & 0.000 \\
&  & 800& &0.068& 0.050& 0.000& &0.075& 0.062 & 0.000 \\
&  & 1200& &0.070& 0.051& 0.001& &0.074& 0.056 & 0.000 \\
\hline\hline
\end{tabular}}
\centering
\end{table}

\begin{figure}[htbp]
\caption{Density plots of $U_{n}/\surd{\mathrm{var}(U_{n})}$ and $N(0, 1)$} 
\label{factorplot}
\begin{center}
\includegraphics[width=0.65\textwidth]{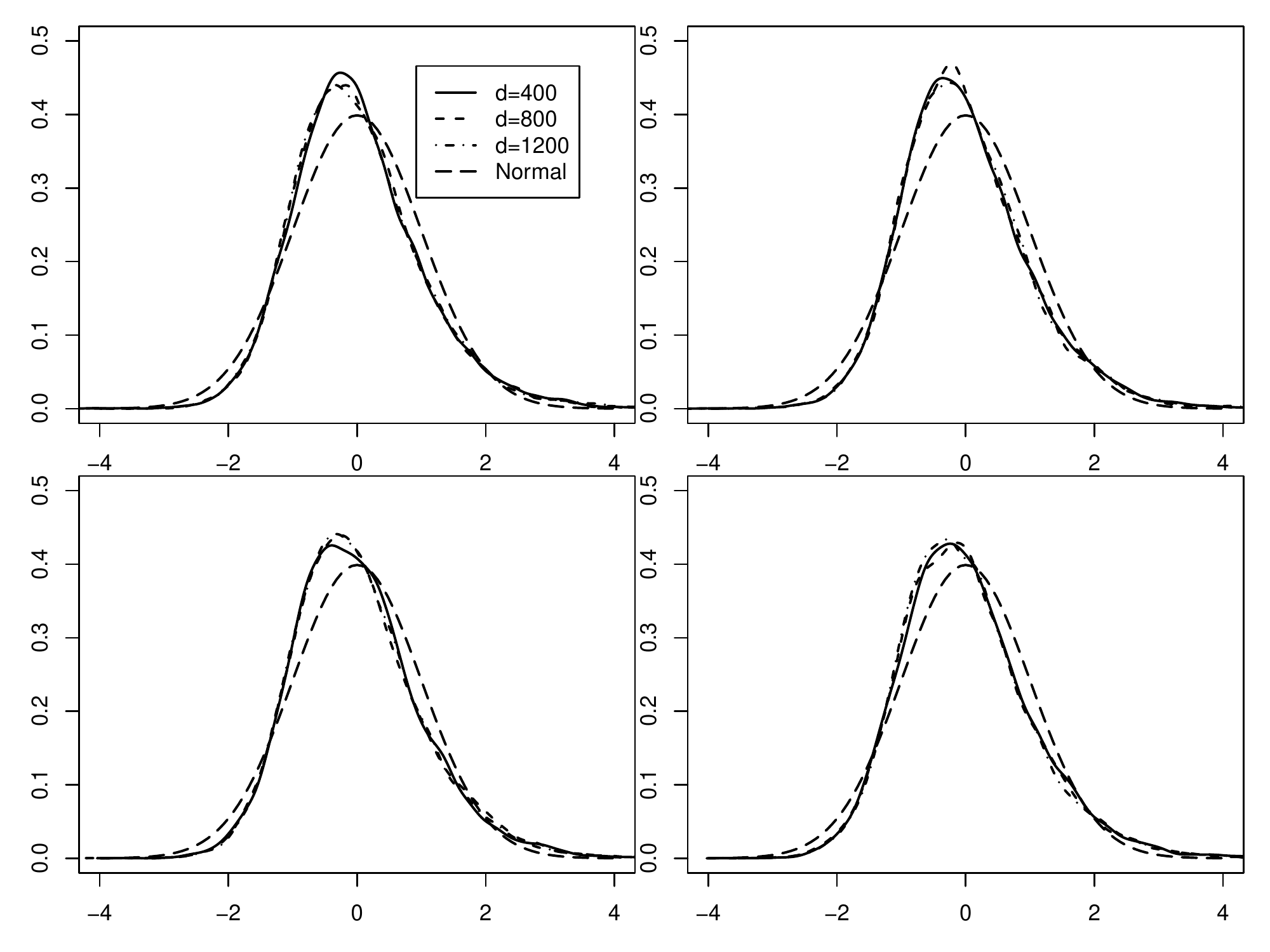}
\end{center}
\end{figure}

For Example~\ref{example1}, the Gaussian multiplier bootstrap method outperforms other two procedures in size accuracy for all cases. The SK test suffers from size distortion. The test by CLT inflates the size more than the GMB method, which can be explained by the fact that condition~\eqref{test_statistic_MANOVA} does not hold and the CLT for $U_{n}$ fails. More specifically, for both $\theta = 0.3$ and $\theta = 0.7$, elementary calculations show that $\lambda_{1}(\Sigma)/\varsigma \to 1$. As a result, \eqref{condclt} is violated as $m = 10$; see also the comment at the end of Section~\ref{secU} for discussion on the non-normality of $U_{n}$. To have more insight, we display in Figure~\ref{factorplot} the density plots of $U_{n}/\surd{\mathrm{var}(U_{n})}$ for $n = 100$ as well as the density of $N(0, 1)$. As we can see from the plots, the distribution of $U_{n}/\surd{\mathrm{var}(U_{n})}$ is skewed to the right for all cases, which explains the inflated sizes of the CLT test.

More simulation studies on power comparison of these three tests are conducted in Section $7.1$.



\section{Data analysis}
\begin{figure}[htbp]
\caption{Histograms of the p-values for testing the association between individual omics feature and the variable of interest after adjusting for the confounder.} 
\label{fig-data}
\begin{center}
\includegraphics[width=0.8\textwidth]{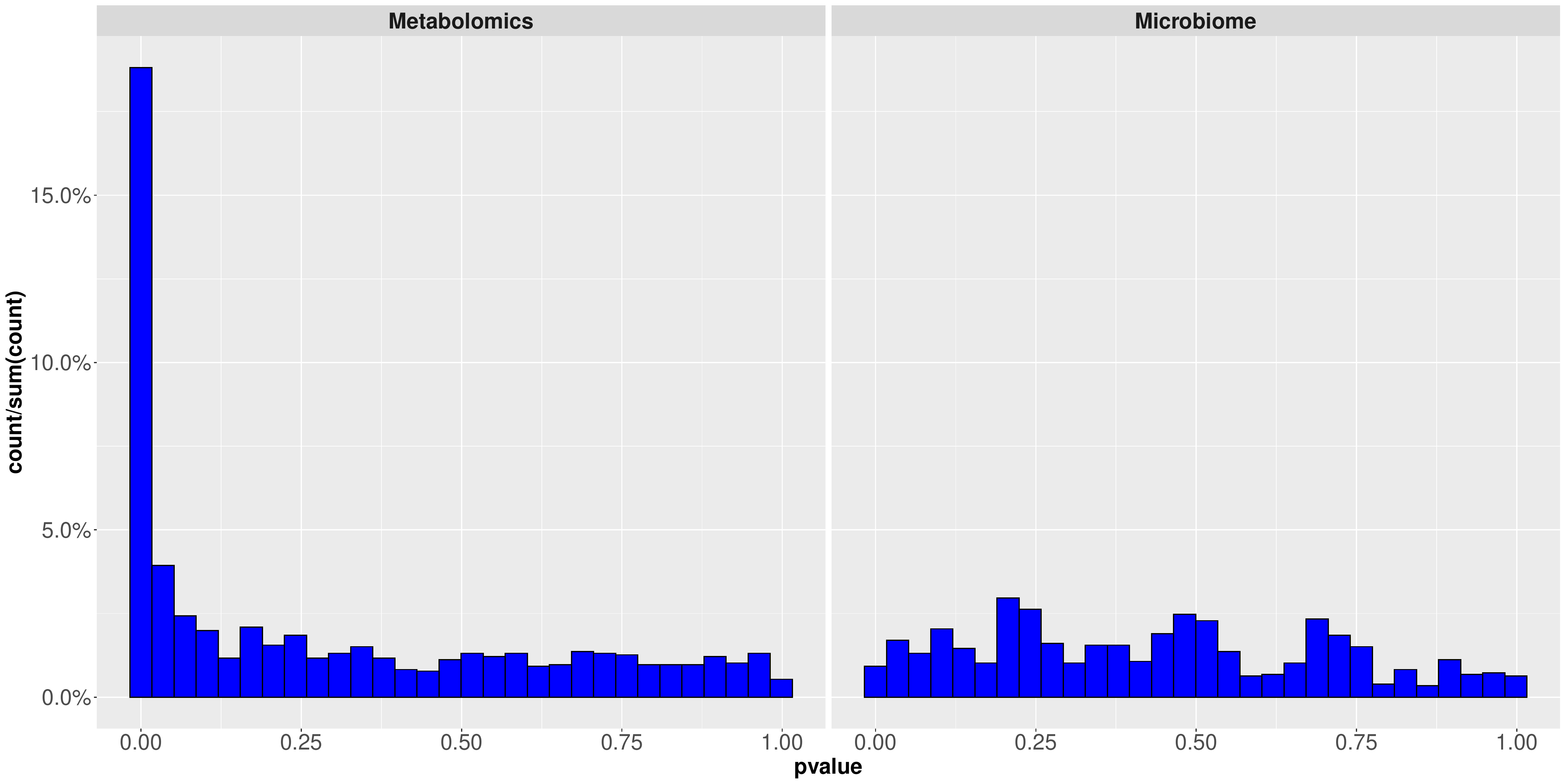}
\end{center}
\end{figure}

We apply the proposed method to two data sets. Our first dataset came from a study of the impact of the gut microbiome on host serum metabolome and insulin sensitivity in non-diabetic Danish adults~\citep{pedersen2016human}. It consists of measurements of 1201 metabolites (325 serum polar metabolites and 876 serum molecular lipids) on 289 serum samples using mass spectrometry. The cleaned dataset was downloaded from https://bitbucket.org/hellekp/clinical-micro-meta-integration~\citep{pedersen2018computational}. We use this data set to identify insulin resistance (IR)-associated metabolites. IR was estimated by the homeostatic model assessment~\citep{pedersen2016human}. Body mass index (BMI) is a confounder for this dataset since it is highly correlated with IR (Spearman’s $\rho = 0.67$) and is known to affect the serum metabolome. Two samples without IR measurement were excluded. For metabolites with zero measurements, zeros were replaced by half of the minimal nonzero value. Log transformation was performed to make the data more symmetrically distributed before analysis. The p-values associated with the three methods (GLT, GMB, and SK) are all very close to zero, indicating a strong dependence between metabolites and IR. We further perform a linear regression analysis on each metabolite using IR and BMI as the covariates. Figure~\ref{fig-data} (left panel) presents the histogram of the p-values on testing the significance of the coefficients associated with IR. We see a high peak close to zero, which provides strong evidence on the association between metabolites and IR. We further apply the Holm–Bonferroni procedure to the p-values to control the family-wise error rate at the 5\% level, resulting in 164 discoveries.

Our second dataset is from the study of the smoking effect on the human upper respiratory tract~\citep{charlson2010disordered}. The original data set contains samples from both throat and nose microbiomes and both body sides. Here we focus on the throat microbiome of the left body side, which includes 60 subjects consisting of 32 nonsmokers and 28 smokers. More precisely, the data set is presented as a $60 \times 856$ abundance table recording the frequencies of detected operational taxonomic units (OTUs) in the samples using the 16S metagenomics approach, together with a metadata table capturing the sample-level information, including the smoking status and sex. We transform the OTU abundance using center log-ratio (CLR) transformation after adding a pseudo-count of 0.5 to the zero counts. Our goal is to test the association of throat microbiomes with smoking status adjusting for sex. The proposed method using either the normal approximation or bootstrap approximation detects a strong association between the throat microbiomes with smoking status. In contrast, the SK method fails to discover the association.

We further perform an OTU-wise linear regression analysis using each OTU (after the CLR transformation) as the response and the smoking status and sex as covariates. Figure~\ref{fig-data} (right panel) presents the histogram of the p-values for testing the association between each OTU and smoking status after adjusting sex in each linear regression. Interestingly, adjusting the multiplicity using either the Holm–Bonferroni procedure or the BH procedure at the $5\%$ level gives zero discovery~\citep{zhou2021linda}. These results suggest that the association between individual OTU and smoking status is weak. However, after aggregating the weak effects from all the OTUs, the combined effect is strong enough to be detected by the proposed method.

\begin{table}[htbp]
\centering
\caption{P-values of the three methods applying to the metabolomics and microbiome data sets.}
\label{pvaluereal}
\scalebox{0.9}{\begin{tabular}{cccccccc}
\hline\hline
&\multicolumn{3}{c}{Metabolomics}&&\multicolumn{3}{c}{Microbiome}\\
\cline{2-4}\cline{6-8}
 &CLT & GMB & SK  & & CLT & GMB & SK\\
\hline
p-value            &0.00 & 0.00 & 0.00 & & $9.7\times 10^{-6}$ & 0.002 & 0.13\\
\hline\hline
\end{tabular}}
\centering
\end{table}

\bibliographystyle{plainnat}
\bibliography{reference}
\end{document}